\documentclass[twocolumn,preprintnumbers,amsmath,amssymb,longbibliography]{revtex4-1}

\usepackage{graphicx}% Include figure files
\usepackage{dcolumn}% Align table columns on decimal point
\usepackage{bm}% bold math
\usepackage{wasysym}% integrals, etc
\usepackage{subfigure}% subfigures
\usepackage{float}% figuras fixadas
\usepackage{tabularx}
\usepackage{color}

\newcolumntype{C}[1]{>{\centering\let\newline\\\arraybackslash\hspace{0pt}}m{#1}} % column type for the tables
\newcolumntype{Y}{>{\centering\arraybackslash}X} % same

\begin{document}

\preprint{}

\title{Tight Binding Parametrization of Few-layer Black Phosphorus from First-Principles Calculations}% Force line breaks with \\

\author{Marcos G. Menezes$^1$}
 \email{marcosgm@if.ufrj.br}
\author{Rodrigo B. Capaz$^{1}$}%
\affiliation{$^1$ Instituto de F\'{i}sica, Universidade Federal do Rio de Janeiro, Caixa Postal 68528 21941-972, Rio de Janeiro, RJ, Brazil}

\date{\today}% It is always \today, today,

\begin{abstract}
\b
We employ a tight-binding parametrization based on the Slater Koster model in order to fit the band structures of single-layer, bilayer and bulk black phosphorus obtained from first-principles calculations. We find that our model, which includes $9$ or $17$ parameters depending on whether overlap is included or not, reproduces quite well the \textit{ab-initio} band structures over a wide energy range, especially the occupied bands. We also find that the Inclusion of overlap parameters improves the quality of the fit for the conduction bands. On the other hand, hopping and on-site energies are consistent throughout the different systems, which is an indication that our model is suitable for calculations on multilayer black phosphorus and more complex situations in which first-principles calculations become prohibitive, such as disordered systems and heterostructures with a large lattice mismatch. We also discuss the limitations of the model and how the fit procedure can be improved for a more accurate description of bands in the vicinity of the Fermi energy.
\end{abstract}

\maketitle

\section{Introduction\label{sec:introduction}}

Since the experimental discovery of graphene in 2004, a strong scientific effort has been employed in the study of this material. Many interesting structural and electronic properties were uncovered, with potential applications in the near future \cite{neto-rmp, novoselov-nature}. Despite its large carrier mobility in comparison with current silicon based devices, graphene is a zero-gap semiconductor, which undermines its application in electronic devices such as field effect transistors, which require a band gap. As such, a lot of research has been done in finding ways to open a band gap in graphene. One of the most promissing routes so far is the application of external electric fields or doping in multilayer systems \cite{ohta-science, expABC, MacDonaldABC, guinea-prb}.

\begin{figure}[ht]
\centering
\includegraphics[width=0.48\textwidth]{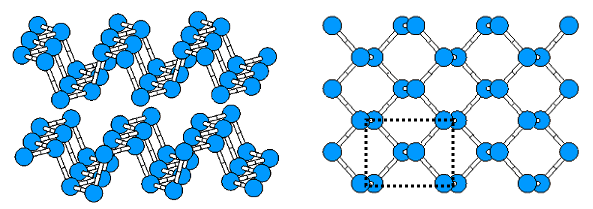}
\caption{\label{fig:BP} Left: Crystal structure of black phosphorus, showing the puckered zigzag chains in each layer. Right: Top view of a single layer, also known as phosphorene. The dashed rectangle indicates an unit cell of the layer.}
\end{figure}

The success of graphene research and the search for a band gap also led to the study of other materials that share some structural similarities to graphene, the so-called 2D materials. These materials form a single layer of thickness of one up to a few atoms, and can be metallic, semiconducting or insulating, thus being suitable for many different applications \cite{ferrari-roadmap, 2D-acsnano}. Moreover, the weak van der Waals interactions between such layers allows the possibility of stacking them in many different ways in order to target specific electronic and optical properties, resulting in a set of materials now known as van der Waals heterostructures \cite{geim-vdw,  novoselov-vdw}. Among such materials, few-layer black phosphorus (BP) is a promissing candidate for future applications. Like graphite, it is a natural layered material that can be exfoliated down to a single layer, which is known as phosphorene in analogy to graphene  \cite{li-fet, lu-raman}. Fig. \ref{fig:BP} shows the crystal structure of bulk and single layer BP. We can see that each layer is composed of puckered zigzag chains of P atoms, a structure that leads to highly anisotropic transport and optical properties \cite{qiao-anisotropy, liu-gap, xia-anisotropy, xiaomu-excitons}. Moreover, theoretical predictions and experimental observations have confirmed that few-layer BP is a semiconductor with a band gap that can be tuned by the number of stacked layers, ranging from $1.73$ eV for single-layer down to $0.35$ eV in the bulk, which is a very desirable property for applications in optoelectronic devices \cite{felipe-natnano2017, tran-gw, liu-gap}.

From a theoretical point of view, the electronic and structural properties of few-layer BP can be accurately determined by first principles calculations, such as those based on Density Functional Theory (DFT) and the GW approximation for quasiparticle corrections \cite{qiao-anisotropy, tran-gw, elahi-prb, zeng-jchem, li-jchem}. However, such methods quickly become prohibitive with an increasing number of atoms in the unit cell, as is the case in calculations for transport properties in disordered systems and heterostructures with a large lattice mismatch. In such situations, semi empirical tight-binding (TB) calculations are often preferred, but they require an adequate parametrization, which may be obtained from theoretical calculations such as DFT and GW, or experimental data. There are a few available parametrizations so far, with different degrees of success in describing the electronic properties of this material. In one class of parametrizations, based on projection techniques using Wannier functions and GW calculations, the proposed models include a single $p_z$-like orbital per atom \cite{rudenko-tb, rudenko-tb2}. Therefore, they do not fully capture the physics of $sp^3$ bonding in this system, due to the staggered nature of each layer. Such models include up to $15$ parameters and are successful in describing the low energy properties of few-layer BP, specially near the $\Gamma$ point of the Brillouin Zone, where the bands are mostly of $p_z$ character. However, they do not fully describe these bands away from the $\Gamma$ point, since they also contain contributions from $s$, $p_x$ and $p_y$ orbitals. Moreover, they describe only a limited number of bands, not taking into account most of the higher energy states. In a second class of parametrizations, based on atomic orbitals, the proposed models include all $s$ and $p$ orbitals, but employ a large number of parameters and can become very complex \cite{paez-tb}. The reason for that is that the hopping between a pair of sites is parametrized individually (considering the symmetries of the structure), so they do not obey an analytical functional form with distance. Therefore, the number of parameters increases with the number of neighboring interactions considered, reaching up to 58 parameters for 8th neighbors. Such models can describe the valence and conduction bands of few-layer BP with a greater overall accuracy than the previous class of models, but they also fail to fully describe the higher energy bands, since the parameters are tailored to fit the low-energy states.

In this work, we provide a simple, yet reliable, tight-binding parametrization based on the Slater-Koster model for the electronic structure of few-layer BP, obtained from first-principles calculations. Our model includes $s$ and $p$ orbitals for each site and an exponential decay behavior for the hopping and overlap integrals, resulting in a number of parameters that range from $9$ up to $17$, depending on whether overlap parameters are included or not. The number of neighboring interactions can be controlled by a cutoff radius, whose value is chosen after convergence tests. The main advantages of this model, besides its simplicity, are that the associated TB parameters have a clear physical interpretation in terms of chemical bonding and that we can fit a larger number of bands in a wide energy range, as we describe below. Our paper is organized as follows: in the next section we present our methodology, including the first-principles calculations used to obtain the band structures, our tight-binding model used to fit them and the fitting procedure. In Sec. \ref{sec:results}, we present our results by discussing calculations with and without overlap integrals, different energy ranges and how the resulting TB parameters compare with previous parametrizations. We also discuss the role of the basis choice in the first-principles calculations and the consistency of our results for single-layer, bilayer and bulk BP. Finally, in Sec. \ref{sec:conclusions} we present our conclusions and discuss the suitability of our model for multilayer systems and more complex situations. We also discuss how the present model can be improved in order to reach a better fit quality.

\section{Methodology \label{sec:methodology}}
\subsection{First-Principles Calculations \label{subsec:dft}}

In the first step of our calculations, we perform first-principles calculations for single-layer, bilayer and bulk BP based on Density Functional Theory (DFT), as implemented on the SIESTA code \cite{ho-kohn, kohn-sham, siesta}. We employ a double-zeta-polarized (DZP) pseudoatomic orbital basis to expand the wavefunctions. We have also perfomed calculations with a simpler single-zeta basis (SZ), which bears a closer correspondence to the $sp^3$ TB model used for the fits. We have found similar results for both basis sets, so we only discuss the results of the DZP basis here. A PBE exchange-correlation functional is used for the electron-electron interactions and norm-conserving Troullier-Martins pseudopotentials are employed for the ion-electron interactions \cite{perdew-burke-ernzerhof, troullier-martins}. The Brillouin Zone is sampled using a $10 \times 10 \times 1$ Monkhorst-Pack k-point grid for single-layer and bilayer BP and a $10 \times 10 \times 4$ for bulk BP \cite{monkhorst-pack}. For single-layer and bilayer BP, a vacuum distance of $20$ \AA \ is used in order to isolate the slabs from their periodic images. 

Since experimental values for single-layer and bilayer BP are not yet available and theoretical predictions present variations depending on the choice of basis, exchange-correlation functional and inclusion of van-der-Waals interactions \cite{qiao-anisotropy, elahi-prb, zeng-jchem, li-jchem}, we fix all structural parameters to the experimental values of bulk BP in all cases \cite{bp-exp}. Finally, since DFT underestimates energy gaps, we perform a rigid shift of all conduction bands in order to reproduce the quasiparticle gaps obtained from GW calculations (a procedure also known as scissors shift) \cite{tran-gw}. It is found in many cases that such a shift is the main effect of the quasiparticle corrections in semiconductors, so it is enough for the purposes of this work \cite{louie-gw}.

\subsection{Tight-Binding Calculations: Slater-Koster Model \label{subsec:sk}}

In our TB model, we use an atomic basis with $3s$, $3p_x$, $3p_y$ and $3p_z$ orbitals for each atom, resulting in a $16 \times 16$ hamiltonian in k-space. Following the prescription of the Slater-Koster model within the two-center approximation, any hopping or overlap integral between orbitals centered at different atomic sites can be expressed in terms of their relative orientations and four integrals related to $\sigma$ and $\pi$ bonding \cite{slater-koster}. 

The simplest case is the hopping between two equivalent $s$ orbitals separated by a displacement vector $\vec{r}$ pointing from one to another. It is given simply by $V_{ss}(\vec{r}) = V_{ss}(r)$ and does not depend on the orientation of the bond. The hopping between a $s$ orbital and a $p_i$ ($i = x, y, z$) orbital separated by a displacement vector $\vec{r}$ from $s$ to $p$ is given by:
\begin{equation}
V_{sp_i}(\vec{r}) = l_i V_{sp\sigma}(r),
\end{equation}
where $l_i = \vec{r}.\hat{x_i}/r$ is the direction cosine of $\vec{r}$ with respect to the corresponding cartesian direction and $V_{sp\sigma}(r)$ is the $\sigma$ hopping between a $s$ orbital and a $p$ orbital directed along the bond. Similarly, the hopping between $p_i$ and $p_j$ orbitals ($i \ne j$) is given by:
\begin{equation}
V_{p_i p_j}(\vec{r}) = l_i l_j [V_{pp\sigma}(r) - V_{pp\pi}(r)],
\end{equation}
where $V_{pp\sigma}(r)$ is the $\sigma$ hopping between $p$ orbitals directed along the bond and $V_{pp\pi}(r)$ is the $\pi$ hopping between $p$ orbitals directed perpendicular to the bond and parallel to each other. For two equivalent $p$ orbitals centered at different sites, the expression is:
\begin{equation}
V_{p_i p_i}(\vec{r}) = l_i^2 V_{pp\sigma}(r) + (1-l_i^2) V_{pp\pi}(r).
\end{equation}
Therefore, all hoppings can be expressed in terms of direction cosines and the amplitudes $V_{ss}(r), V_{sp\sigma}(r), V_{pp\sigma}(r)$ and $V_{pp\pi}(r)$ calculated at the bond length. These amplitudes decrease as the distance between the orbitals increase. Since the atomic wavefunctions decay exponentially for large distances away from the nuclei, we model the hopping amplitudes in the same way:
\begin{equation}
V_{t}(r) = \left\{ \begin{array}{lr}
            V_{t}(r_0) \exp \left( \frac{r-r_0}{r_d^t} \right), & \text{for } r < r_{cut} \\
            0, & \text{for } r \ge r_{cut}
            \end{array} \right.
\end{equation}
In the above equation, $t = ss, sp\sigma, pp\sigma, pp\pi$ is an index that labels the type of hopping. $r_0$ is a reference distance at which the hopping amplitudes are calculated, which we choose as the nearest neighbor distance: $r_0 = 2.224$ \AA. $r_d$ is a decay distance and $r_{cut}$ is a cutoff distance. We set $r_{cut}$ to $10.0$ \AA, which generates well converged band structures according to our tests.

With this model, we have $8$ parameters related to hopping: four amplitudes $V_{ss}$, $V_{sp\sigma}$, $V_{pp\sigma}$ and $V_{pp\pi}$ and four decay distances $r_d^{ss}$, $r_d^{sp\sigma}$, $r_d^{pp\sigma}$, $r_d^{pp\pi}$. By setting the on-site energy of the $p$ orbitals to zero and including that of the $s$ orbitals in the model ($E_s$), we end up with a minimal model with a total of $9$ adjustable parameters. We may also include overlap parameters following the same prescription outlined above, by adding up to four corresponding amplitudes and decay distances, resulting in a model with a maximum of $17$ adjustable parameters. In the next section, we present results for both the minimum model and the model with all overlap parameters.

\subsection{Parametrization: Least-Squares Minimization}

In order to fit the Tight-Binding band structures to the corresponding DFT results, we use a least-squares minimization. In this method, we consider the TB band structure as a function of the Slater-Koster parameters described above and minimize the following $\chi^2$ function:
\begin{equation}
\label{eq:chi2}
\chi^2 = \frac{1}{N_{data}} \sum_{n, \vec{k}} \left[ \left( E_{n, \vec{k}}^{TB} - E_F^{TB} \right) - \left( E_{n, \vec{k}}^{DFT} - E_F^{DFT} \right) \right]^2,
\end{equation}
where $E_{n,\vec{k}}^{DFT}$ and $E_{n,\vec{k}}^{TB}$ are the eigenvalues and $E_F^{DFT}$ and $E_F^{TB}$ are the corresponding Fermi energies from DFT and TB calculations, respectively. The summation in this equation can be restricted to a given set of bands, k-points or energy range, so the number of data points $N_{data}$ can be different in each case. For a full-adjustment, in which all bands and k-points are included, $N_{data} = N_{bands} \times N_k$. We have considered different sets of k-points and energy ranges, as we discuss in the next section.

The minimization of $\chi^2$ was carried on using Powell's minimization algorithm \cite{num-rec}. The iterative procedure stops when the relative value of $\chi^2$ differs by less than $10^{-4}$ between two consecutive iterations. We find in our calculations that the minimized value of this function depends on the initial choice of parameters, especially when overlap is included. This is an indication that this functions may contain different local minima, so we have tested several different starting points and looked for the best fit.

\section{Results \label{sec:results}}

\subsection{Full Adjustment}

\begin{figure*}[ht]
\centering
\includegraphics[width=0.8\textwidth]{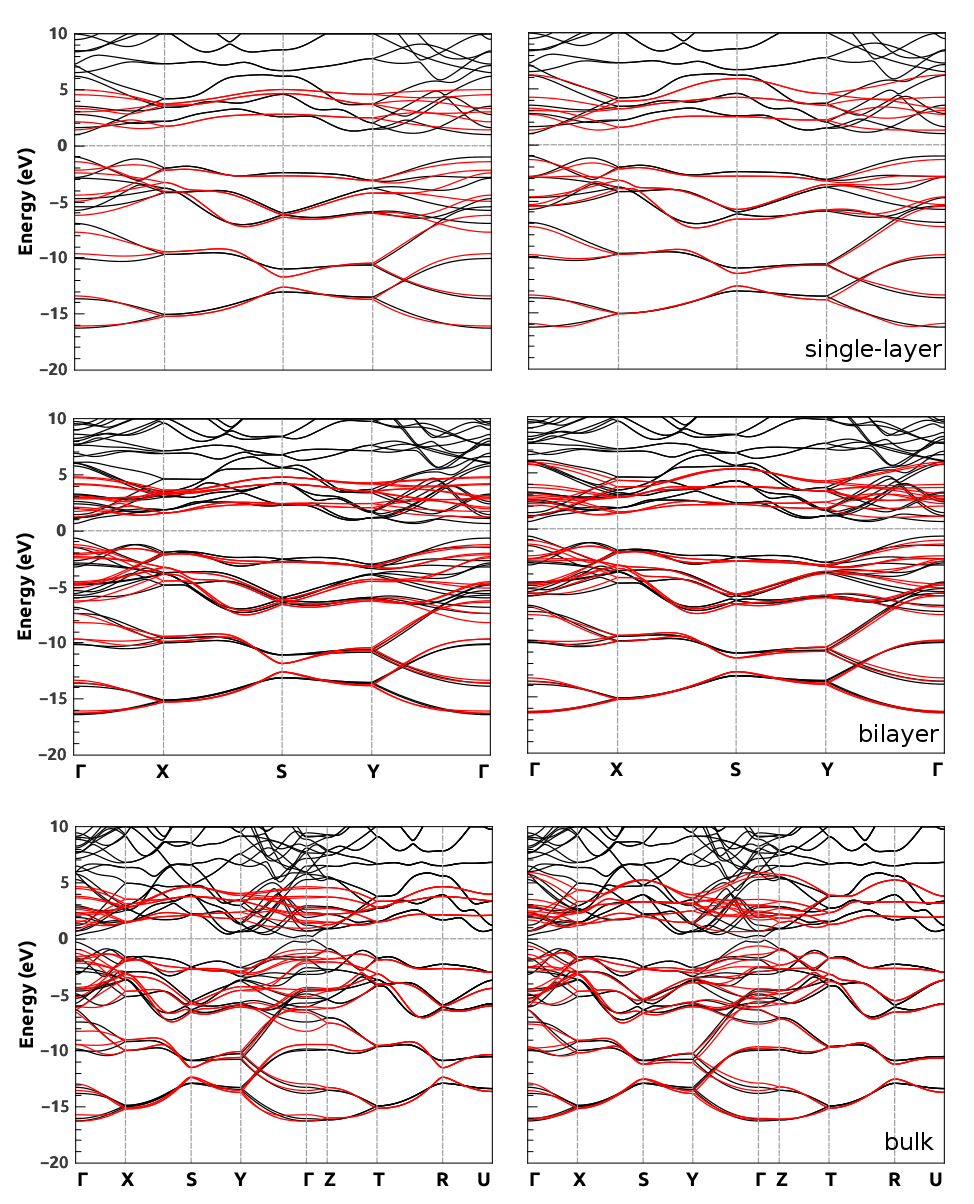}
\caption{\label{fig:full_adj} Comparison between DFT band structures (black) and corresponding Tight-Binding fits (red) for single-layer (top), bilayer (middle) and bulk Black Phosphorus (bottom). The left (right) panels correspond to a model without (with) overlap parameters. The Fermi energy is set to zero in all cases (horizontal dashed line).}
\end{figure*}

In Fig. \ref{fig:full_adj}, we show a comparison between the DFT band structures of single-layer, bilayer and bulk BP and the corresponding TB full adjustments without (left) and with (right) overlap parameters in the model. For bulk BP, we have used a conventional rectangular unit cell with four atoms in order to facilitate the comparison with the single-layer and bilayer Brillouin Zones. As we can see, the minimum TB model without overlap parameters gives a good overall description of the DFT band structure, especially the occupied bands. The inclusion of overlap parameters in the model improves the description of the unoccupied band and fixes a few avoided crossings, leading to a better quality fit.

Table \ref{tab:full_adj_param} shows the fit parameters for each case and the corresponding standard deviation of the adjusment, which is the minimized value of $\sigma = \sqrt{\chi^2}$. The first three (next three) rows of the table correspond to a model without (with) overlap parameters, which we label TB1 (TB2). The last row contains reference parameters from Harrison's model, which describes the hopping between $s$ and $p$ orbitals as a $1/r^2$ decay \cite{harrison-book}. We can see that the hopping parameters are consistent throughout single-layer, bilayer and bulk BP, which is an indication that this model is suitable for calculations on multilayer BP and more complex situations. When overlap parameters are included in the model, the adjusted values for the hopping amplitudes and decay constants change, but they remain consistent throughout the three cases. Overlap amplitudes are found to be very small, with the exception of $S_{pp\pi}$, which may be an indication that the fit procedure is less sensitive to these parameters. Therefore, a model with all hopping parameters and only $S_{pp\pi}$ and $r_{pp\pi}^{ds}$ might give a good description of the electronic structure.

\begin{table*}[t]
\centering
\caption{Tight-Binding paramters for a full adjustment of the DFT band structure, as shown in Fig. \ref{fig:full_adj}. Hopping and overlap amplitudes are calculated at nearest neighbor distance ($r_0 = 2.224$ \AA) and are in eV. Decay distances are in \AA \ and $\sigma$ is the standard deviation of the adjusment in eV. The last line contains reference parameters from Harrisons' model \cite{harrison-book}. Zero values mean that they are smaller than $0.01$ in the corresponding units.} 
\label{tab:full_adj_param}
\begin{tabularx}{\textwidth}{X *{18}{Y} }
$N_{layers}$ & $V_{ss}$ & $V_{sp\sigma}$ & $V _{pp\sigma}$ & $V_{pp\pi}$ & $E_{s}$ & $r_{ss}^d$ & $r_{sp\sigma}^d$ & $r _{pp\sigma}^d$ & $r_{pp\pi}^d$ & $S_{ss}$ & $S_{sp\sigma}$ & $S _{pp\sigma}$ & $S_{pp\pi}$ & $r_{ss}^{ds}$ & $r_{sp\sigma}^{ds}$ & $r _{pp\sigma}^{ds}$ & $r_{pp\pi}^{ds}$ & $\sigma$ \\
\hline
\multicolumn{19}{c}{Without Overlap (TB1)} \\
\hline
1    & -1.59 & 2.39 & 4.03 & -1.14  & -8.80 & 0.33 & 0.53 & 0.58 & 0.53 & -         & -         & -        & -        & -         & -        & -         & -        & 0.54 \\
2    & -1.56 & 2.43 & 3.91 & -1.18 & -8.59 & 0.36 & 0.54 & 0.59 & 0.44 & -         & -         & -        & -        & -         & -        & -         & -        & 0.55 \\
Bulk & -1.54 & 2.37 & 3.63 & -1.25 & -8.44 & 0.35 & 0.59 & 0.58 & 0.41  & -         & -         & -        & -        & -         & -        & -         & -        & 0.58 \\
\hline
\multicolumn{19}{c}{With Overlap (TB2)} \\
\hline
1    & -2.02 & 2.76 & 3.99 & -1.07 & -7.34 & 0.43  & 0.55 & 0.76 & 0.51 & 0.06 & 0.01 & 0.04 & 0.14 & 0.04  & 6.37 & 0.00 & 0.36 & 0.45 \\
2    & -2.01  & 2.63 & 3.91 & -1.09 & -8.13 & 0.48 & 0.53 & 0.71 & 0.36 & $ 0.05 $ & 0.01 & 0.00 & 0.12 & 0.02 & 0.20 & 0.00 & 0.30  & 0.47 \\
Bulk & -1.75 & 2.75 & 3.61 & -1.07 & -7.38  & 0.48 & 0.56 & 0.73 & 0.35 & 0.04  & 0.01  & 0.00 & 0.15 & $ 0.01 $ & 0.00 & 0.00 & 0.40  & 0.50  \\
Har & -2.03 & 2.19 & 3.42 & -0.97 & -9.68 & - & - & - & - & - & - & - & - & - & - & - & - & -
\end{tabularx}
\end{table*}
\begin{table*}[ht]
\centering
\caption{Electron ($m_e$) and hole ($m_h$) effective masses along $\Gamma-X$ (x) and $\Gamma-Y$ (y) and energy gap at the $\Gamma$ point ($E_g$) from DFT and TB fits. TB1 is the model which does not include overlap parameters, while the TB2 model contains all parameters. Theoretical reference values are taken from Ref. \cite{qiao-anisotropy}.  Masses are in units of the bare electron mass and gap energies are in eV.}
\label{tab:full_adj_em}
%\begin{tabular}{| l | *{4}{C{0.90cm}} | *{4}{C{0.90cm}} | *{3}{C{0.90cm}} | }
\begin{tabularx}{\textwidth}{ X | *{4}{Y} | *{4}{Y} | *{3}{Y} }
\hline
		&     \multicolumn{4}{ c |}{ single-layer }   & \multicolumn{4}{ c |}{ bilayer } & \multicolumn{3}{ c }{ bulk } \\
\cline{2-12}
                   & DFT & TB1 & TB2  &  Ref.  & DFT & TB1  & TB2  & Ref. & DFT  & TB1 & TB2 \\
\hline
$m_h^x$ & 0.13 & 0.40 & 0.37 & 0.15 & 0.10 & 0.37 & 0.32 & 0.15 & 0.15 & 0.34 & 0.26 \\
$m_e^x$ & 0.13 & 0.44 & 0.40 & 0.17 & 0.10 & 0.44 & 0.32 & 0.18 & 0.17 & 0.37 & 0.25 \\
$m_h^y$ & 3.42 & 2.37 & 1.58 & 6.35 & 1.71 & 1.90 & 1.35 & 1.81 &  -       & -        & - \\       
$m_e^y$ & 1.14 & 2.37 & -        & 1.12 & 1.14 & 4.73 & -        & 1.13 &  -       & -        & - \\
 $E_g$      & 2.00 & 2.83 & 2.62 & 1.53 & 1.30 & 2.48 & 2.12 & 1.00 & 0.59 & 1.88 & 1.29 \\
 \hline
\end{tabularx}
\end{table*}

For transport properties, the most relevant bands are those in the vicinity of the Fermi energy. Therefore, if the model is intended to be used for such applications it is desirable that the fit reproduces well these bands, especially for k-points in the vicinity of the band-edge states. In order to further check the description of these bands, we have calculated the band gaps and effective masses along $\Gamma-X$ and $\Gamma-Y$ for electrons and holes for both DFT and TB calculations. The results are reported in Table \ref{tab:full_adj_em} together with reference values from DFT calculations from Ref. \cite{qiao-anisotropy}. We recall that the unoccupied bands from DFT were shifted in order to match the quasiparticle gap obtained in the reference GW calculation \cite{tran-gw}. In particular, our DFT calculations predicts bulk-BP to be a semimetal, so the shift is specially important in this case. With the energy shift, bulk-BP has an indirect gap with the top of the valence band at the $\Gamma-Z$ line and the bottom of the conduction band at the $\Gamma-Y$ line, so we do not report the corresponding effective masses. Similarly, we can see that the TB model with overlap gives a very flat conduction band at the $\Gamma-Y$ line in all cases, so we do not report the corresponding effective mass. As we can see, both TB models overestimate the gaps and effective masses in all cases. Nevertheless, we see that the full adjustment provides a good description of the whole band structure, with consistent effective masses for single-layer, bilayer and bulk BP and standard deviations $0.45 - 0.60$ eV in an energy range of about $25$ eV. The fit also captures the high effective mass anisotropy along $\Gamma-X$ and $\Gamma-Y$ lines, but at a qualitative level.

\subsection{Adjustments for Low Energy Bands}

We now discuss how the fit procedure can be improved in order to get a better description of the band structure near the Fermi energy. For that end, we focus our attention only on single-layer BP, since the model has already shown to be consistent for multilayer systems. As we are mainly interested in an accurate description of the energy gaps and effective masses near the $\Gamma$ point, we narrow down the energy range and k-point grid of the adjustment, which reduces the number of data points included in the summation in Eq. \ref{eq:chi2}.

\begin{figure}[h]
\centering
\includegraphics[width=0.5\textwidth]{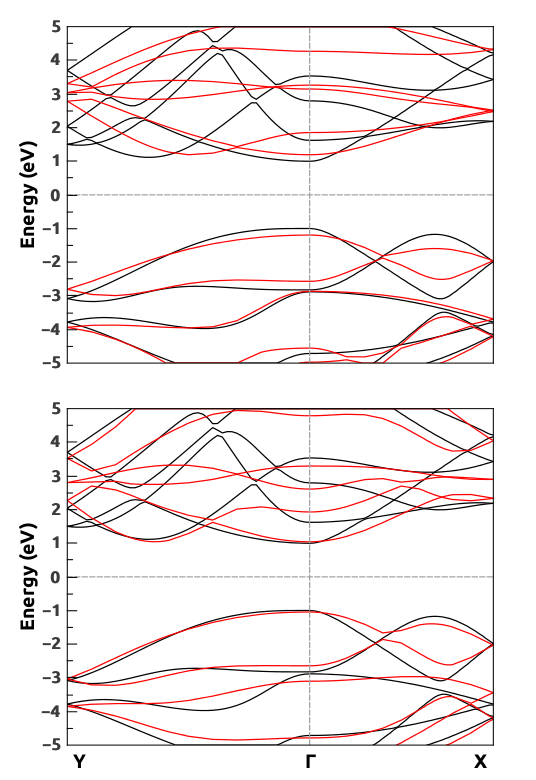}
\caption{\label{fig:ygx_adj} Comparison between DFT (black) and Tight-Binding band structures (red) for single-layer Black Phosphorus for bands in the vicinity of the Fermi Energy, which is set to zero. The top (bottom) panels correspond to a model without (with) overlap parameters.}
\end{figure}

For this fit, we have considered k-points only along the $Y-\Gamma-X$ path and included energy levels only within a $5$ eV window from the Fermi energy. Moreover, since the most relevant features of the valence (conduction) band are near the top (bottom) at the $\Gamma$ point and local maxima (minima) at nearby points, we have sampled these regions with a higher density of k-points. The TB fits for the models with (TB3) and without (TB4) overlap parameters are shown in Fig. \ref{fig:ygx_adj}. As we can see, the full model improves the fit of the valence and conduction bands, at the expense of the deeper occupied bands. The corresponding adjusted parameters are shown in Table \ref{tab:ygx_adj_param}. Note that some of these parameters are very different from those reported in Table \ref{tab:full_adj_param}, especially those involving hoppings and overlaps to $s$ orbitals. In fact, such parameters are more relevant to the description of deeper energy bands, not included in this fit. As such, the fit procedure becomes less sensitive to these parameters and greatly depends on their initial choice. This problem is particularly bad in the full model, where the calculation usually does not converge for energy ranges smaller than $5$ eV because the overlap parameters assume unphysical values during the minimization.

\begin{table*}[t]
\centering
\caption{Tight-Binding parameters for the adjustments shown in Fig. \ref{fig:ygx_adj}. Hopping and overlap amplitudes are calculated at nearest neighbor distance ($r_0 = 2.224$ \AA) and are in eV. Decay distances are in \AA \ and $\sigma$ is the standard deviation of the adjusment in eV.}
\label{tab:ygx_adj_param}
\begin{tabularx}{\textwidth}{X *{18}{Y} }
Model & $V_{ss}$ & $V_{sp\sigma}$ & $V _{pp\sigma}$ & $V_{pp\pi}$ & $E_{s}$ & $r_{ss}^d$ & $r_{sp\sigma}^d$ & $r _{pp\sigma}^d$ & $r_{pp\pi}^d$ & $S_{ss}$ & $S_{sp\sigma}$ & $S _{pp\sigma}$ & $S_{pp\pi}$ & $r_{ss}^{ds}$ & $r_{sp\sigma}^{ds}$ & $r _{pp\sigma}^{ds}$ & $r_{pp\pi}^{ds}$ & $\sigma$ \\
\hline \\
TB3 & -3.58 & 4.39 & 3.78 & -1.43 & -7.34 & 0.22 & 0.54 & 0.82 & 0.68 & -         & -         & -        & -        & -         & -        & -         & -        & 0.33 \\
TB4 & -2.78 & 3.93 & 3.97 & -1.07 & -7.88 & 0.42 & 0.48 & 0.88 & 0.53 & 0.32  & -0.05  & 0.17 & 0.06 & 1.03 & 5.18 & 0.21 & 2.27 & 0.28 \\
\end{tabularx}
\end{table*}

\begin{figure}[h]
\centering
\includegraphics[width=0.5\textwidth]{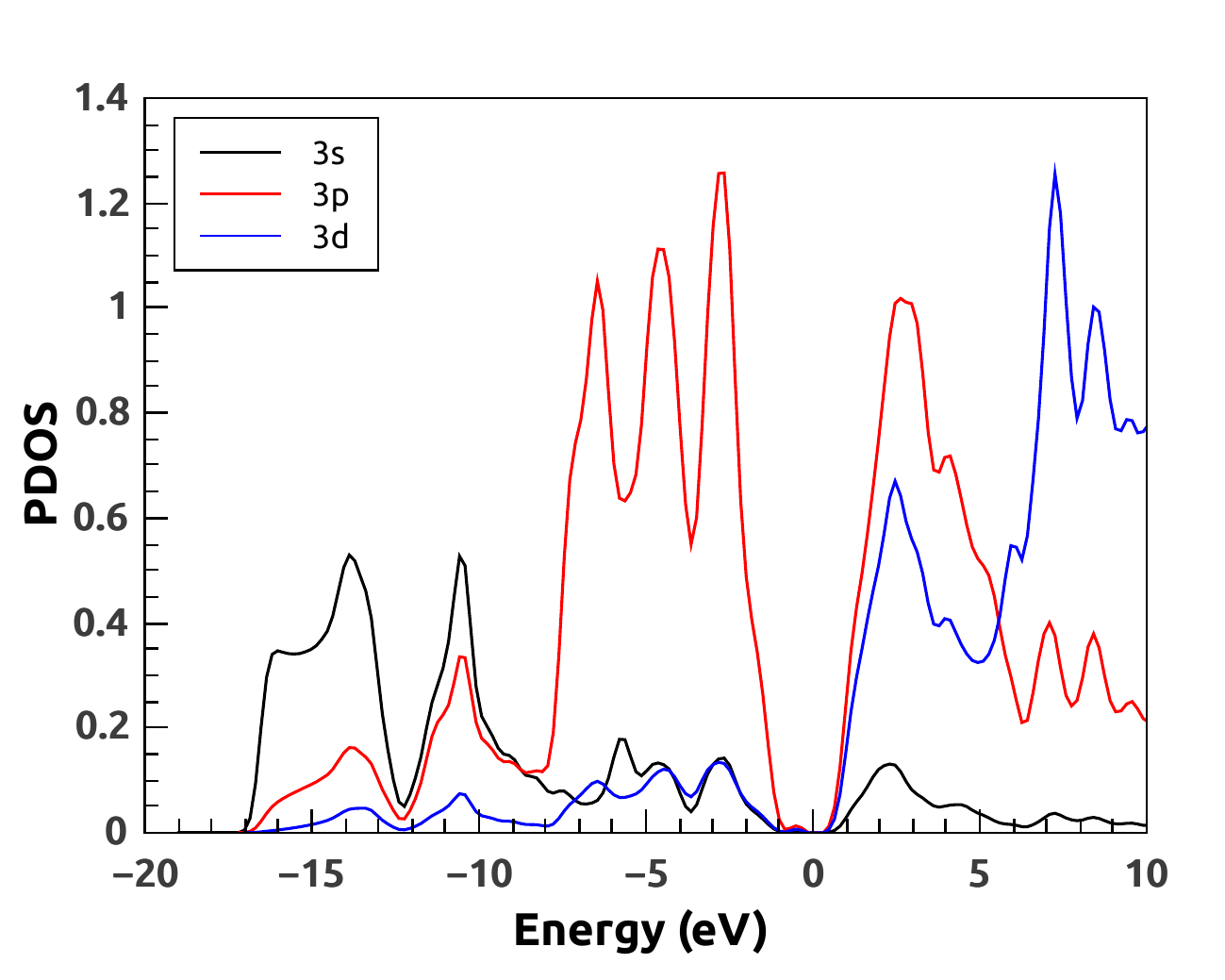}
\caption{\label{fig:pdos} Projected density of states (PDOS) of single-layer BP as given by DFT calculations with the DZP basis. The energy range is the same as the band structures in Fig. \ref{fig:full_adj} and the Fermi energy is set to zero. We have also performed a scissors shift in order to reproduce the correct quasiparticle gap}
\end{figure}

Another feature observed in our calculations is that both models fit better the occupied states, as is evident in Figs. \ref{fig:full_adj} and \ref{fig:ygx_adj}. In fact, the fit of these states can be further improved by completely leaving the unnocupied states out of the adjustment, as verified by our test calculations. On the other hand, even if the occupied states are completely left out of the fit, the adjustment of the conduction bands does not improve. In order to further understand this issue, we have performed projected density of states (PDOS) calculations within DFT. The results for the single-layer are shown in Fig. \ref{fig:pdos}. Bilayer and bulk BP yield similar results, so we do not discuss them here. As we can see, the $d$ orbitals give an important contribution to the conduction band states. Since these orbitals are not included in our TB model, it is expected that the fit quality is not the same for these bands. However, it should also be pointed out that DFT calculations with the SZ basis, which does not include $d$ orbitals, give similar results when we use the same TB model to fit the resulting band structures, even tough the standard deviations are slightly smaller. This may be an indication that our simple model has a limitation in the description of the conduction bands, as also seen in similar models for other semiconductors \cite{vogl-tb}. Another important feature of the PDOS is that the $s$ orbitals contribute mainly to the deeper bands, in agreement with our discussion above in terms of the TB parameters. There is some hybridization with the $p$ orbitals in the low energy range, but the latter still dominate.  Moreover, our calculations indicate that the low-energy bands do not have a full $p_z$ character. Therefore, TB parametrizations which include only a single $p_z$-like orbital per site are expected to not fully capture all features of these bands, as seen in the proposed models so far \cite{rudenko-tb, rudenko-tb2}. However, they are still very successful in fitting these bands in the vicinity of the $\Gamma$ point (and some higher energy states as well), thus correctly reproducing band gaps and effective masses.

\begin{figure}[h]
\centering
\includegraphics[width=0.5\textwidth]{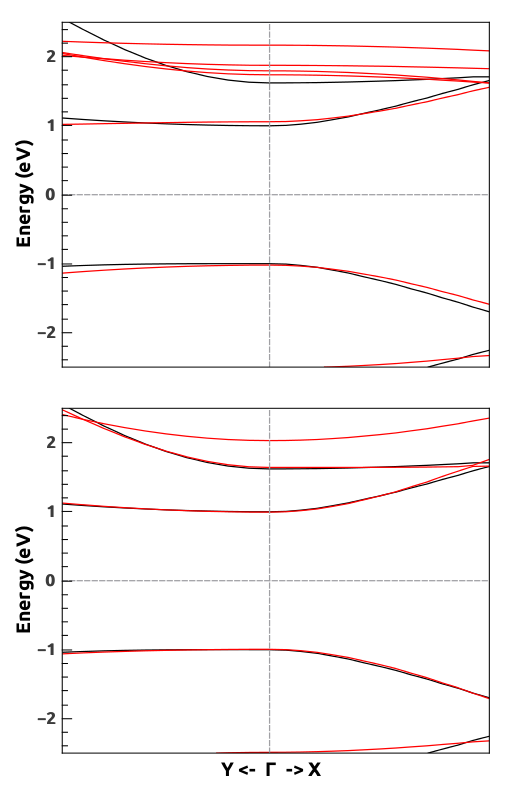}
\caption{\label{fig:narrow_adj} Comparison between DFT (black) and TB band structures (red) for single-layer black phosphorus for bands in a narrow vicinity of the $\Gamma$ point. The top (bottom) panels correspond to a model without (with) overlap parameters.}
\end{figure}

Even with the limitations of our model, we may also get an improved description of the gap and effective masses near the $\Gamma$ point. To this end, we perform a new fit with k-points in a narrow vicinity of the $\Gamma$ point and a $2.5$ eV energy range. Only k-points with $|\vec{k}| < 0.10 \ 2\pi/a$ are included, where $a = 3.3136$ \AA \ is the experimental lattice constant of bulk BP. The results are shown in Fig. \ref{fig:narrow_adj} and the corresponding parameters are reported in Table \ref{tab:narrow_adj_param} (TB5 and TB6 without and with overlap, respectively). Note that the standard deviation of the adjustment is now greatly reduced, ranging from $0.02$ to $0.08$ eV, due to the reduced number of data points. For the same reason, the adjusted parameters are much more sensitive to the fit conditions than those of the full adjustment reported in Table \ref{tab:full_adj_param}.

\begin{table*}[ht]
\caption{Tight-Binding parameters for the adjustments shown in Fig. \ref{fig:narrow_adj}. Hopping and overlap amplitudes are calculated at nearest neighbor distance ($r_0 = 2.224$ \AA) and are in eV. Decay distances are in \AA \ and $\sigma$ is the standard deviation of the adjusment in eV.}
\label{tab:narrow_adj_param}
\begin{tabularx}{\textwidth}{ X *{18}{Y} }
Model & $V_{ss}$ & $V_{sp\sigma}$ & $V _{pp\sigma}$ & $V_{pp\pi}$ & $E_{s}$ & $r_{ss}^d$ & $r_{sp\sigma}^d$ & $r _{pp\sigma}^d$ & $r_{pp\pi}^d$ & $S_{ss}$ & $S_{sp\sigma}$ & $S _{pp\sigma}$ & $S_{pp\pi}$ & $r_{ss}^{ds}$ & $r_{sp\sigma}^{ds}$ & $r _{pp\sigma}^{ds}$ & $r_{pp\pi}^{ds}$ & $\sigma$ \\
\hline \\
TB5 & -0.78 &  3.61 & 3.46 & -1.32 &  -16.4  & 2.15  & 0.36 & 0.49 & 0.01 & - & - & - & - & - & - & - & - & 0.08 \\
TB6 & -2.13 &  2.64 & 4.76 & -1.51 & -10.8 & 0.84 & 0.45  & 0.49 & 0.60 & 0.19 & -0.07 & 0.05 & 0.14 & 0.03 & 0.27 & 0.23 & 0.18 & 0.02 \\
\end{tabularx}
\end{table*}

The effective masses and the gap energies at the $\Gamma$ point are shown in Table \ref{tab:narrow_adj_em}. We can see that the narrow fit greatly improves the description of these quantities in comparison with the full adjustment shown in Table \ref{tab:full_adj_em}, especially the gap and the effective masses along $\Gamma-X$. For the $\Gamma-Y$ direction, the TB model without overlap parameters gives a conduction band minimum outside $\Gamma$, so $m_e^y$ is not reported. On the other hand, the model with overlap gives a good description of both masses. We point out, however, that since the valence band is almost flat along this line, numerical values for the corresponding effective mass can be imprecise.
\begin{table}[h]
\centering
\caption{Electron ($m_e$) and hole ($m_h$) effective masses along $\Gamma-X$ (x) and $\Gamma-Y$ (y) and energy gap at the $\Gamma$ point ($E_g$) from DFT and TB fits in a narrow vicinity of the $\Gamma$ point. TB5 is the model which does not include overlap parameters, while TB6 contains all parameters. Masses are in units of the bare electron mass and gap energies are in eV.}
\label{tab:narrow_adj_em}
\begin{tabularx}{\linewidth}{ X  *{4}{Y} }
                   & DFT  & TB5  & TB6 \\
\hline \\
$m_h^x$ & 0.13 & 0.21 & 0.16 \\
$m_e^x$ & 0.13 & 0.22 & 0.15 \\
$m_h^y$ & 3.42 & 1.18 & 2.68  \\       
$m_e^y$ & 1.14 & -        & 1.07  \\
 $E_g$      & 2.00 & 2.08 & 1.99  \\
\end{tabularx}
\end{table}

\section{Conclusions \label{sec:conclusions}}

In summary, we have developed a Tight-Binding parametrization based on the Slater Koster model in order to describe the band structures of multilayer black phosphorus systems as obtained from first principles calculations. The model includes $s$ and $p$ orbitals and we model the behavior of the hopping and overlap amplitudes with distance as an exponential decay, which reduces the total number of parameters. We find that our model can fit the DFT band structures over an energy range of about $25$ eV with standard deviations of $0.45 - 0.60$ eV. It fits especially well the occupied bands, but we find it has limitations in the description of the conduction bands, even tough the inclusion of overlap parameters improves their fit. The adjusted hopping parameters are consistent throughout single-layer, bilayer and bulk BP, which is an indication the model is suitable for multilayer systems. On the other hand, overlap parameters are in general less consistent and more dependent on the conditions of the fit.

For the valence and conduction bands, we find that the full adjustment, in which all bands are included, does not give a sufficiently accurate description of the energy gaps and effective masses, but it captures the mass asymmetry at a qualitative level. For an improved description of these quantities, the fit procedure needs to be carried on bands in a narrow energy range and k-points in the vicinity of the gap, leading to standard deviations of $0.02-0.08$ eV.

The description of the conduction bands may be improved with the inclusion of fictitious orbitals which lead to band repulsion, as is usually done in TB models for semiconductors \cite{vogl-tb}. The dependence of the hopping and overlap amplitudes with distance can also be modified to more complex expressions, such as the product of polynomials and exponentials. One can also go beyond the two center approximation and include more types of hoppings and overlaps. Finally, we point out that for the study of impurities or heterostructures, local modifications of hopping and on-site energies may also be required. For the least-squares fit, the use of methods that perform a search over all the parameter space, such as genetic algorithms, may lead to improved adjustments. All these modifications will lead to more complex models with an increased number of parameters and longer computation times, so we believe our model provides an important, yet simple starting point for tackling these situations.

\vfill

\begin{acknowledgments}
This work was supported by the following Brazilian funding agencies: CNPq, CAPES, FAPERJ and INCT-Nanomateriais de Carbono. We also thank the computational support from DIMAT-INMETRO.
\end{acknowledgments}

\bibliography{Artigo}% Produces the bibliography via BibTeX.

\end{document}